\documentclass{emulateapj}
\usepackage[greek,english]{babel}
\usepackage[latin9]{inputenc}
\usepackage{amssymb}
\usepackage{textcomp}
\usepackage{natbib}
\usepackage{rotfloat}
\usepackage[T1]{fontenc}
\usepackage{graphicx}
\usepackage{esint}
\usepackage{verbatim}
\usepackage{epstopdf}

\shorttitle{Gas and Dust Absorption in the DoAr 24E System}
\shortauthors{Kruger et al.}

\begin{document}

\title{Gas and Dust Absorption in the DoAr 24E System}

\author{Andrew J. Kruger}
\affil{Department of Physical Science, Wilbur Wright College, 4300 N. Narragansett Ave, Chicago, IL 60634}

\author{Matthew J. Richter}
\affil{Department of Physics, University of California at Davis, One Shields Ave, Davis, CA 95616, USA}

\author{Andreas Seifahrt}
\affil{Department of Astronomy and Astrophysics, University of Chicago, 5640 S. Ellis Ave, Chicago, IL 60637, USA}

\author{John S. Carr}
\affil{Remote Sensing Division, Naval Research Laboratory, Code 7210, Washington, DC 20375, USA}

\author{Joan R. Najita}
\affil{National Optical Astronomy Observatory, Tucson, AZ 85719, USA }

\author{Margaret M. Moerchen}
\affil{European Southern Observatory, Alonso de C\'{o}rdova 3107, Santiago, Chile}
\affil{Leiden Observatory, PO Box 9513, 2300 RA Leiden, The Netherlands}

\and
\author{Greg W. Doppmann}
\affil{W.M. Keck Observatory, 65-1120 Mamalahoa Hwy, Kamuela HI 96743}

\begin{abstract}

We present findings for DoAr 24E, a binary system that includes a classical infrared companion.  We observed the DoAr 24E system with the {\it Spitzer} Infrared Spectrograph (IRS), with high-resolution, near-infrared spectroscopy of CO vibrational transitions, and with mid-infrared imaging.  The source of high extinction toward infrared companions has been an item of continuing interest.  Here we investigate the disk structure of DoAr 24E using the column densities, temperature, and velocity profiles of two CO absorption features seen toward DoAr 24Eb.  We model the SEDs found using T-ReCS imaging, and investigate the likely sources of extinction toward DoAr 24Eb.  We find the lack of silicate absorption and small CO column density toward DoAr 24Eb suggest the mid-infrared continuum is not as extinguished as the near-infrared, possibly due to the mid-infrared originating from an extended region.  This, along with the velocity profile of the CO absorption, suggests the source of high extinction is likely due to a disk or disk wind associated with DoAr 24Eb.

\end{abstract}

\keywords{circumstellar matter -- protoplanetary disks -- stars: pre-main-sequence -- stars: individual (DoAr 24E)}

\section{Introduction}

Infrared Companions (IRCs; Koresko et al.~1997) are stellar objects characterized as being in binary systems with young stars, being faint in the visible, and dominating the system flux at infrared and longer wavelengths.  Being companions to young T Tauri stars suggests they are young stars as well, but the source of high extinction has remained a mystery.  Koresko et al.~(1997) noted the seven classical IRCs because they are well-studied systems, so it is important to identify the sources of extinction for future studies.  As an example, if the extinction is due to IRCs having disks seen edge-on, these systems would present a good opportunity to search for molecular absorption originating in the inner disk regions.  GV Tau N, a classical IRC, was found to exhibit warm molecular C$_2$H$_2$ and HCN absorption (Doppmann et al.~2008; cf. Gibb et al.~2007), rare features previously found in only one other disk system (IRS 46; Lahuis et al.~2006), which provided an important benchmark for inner disk chemistry models (e.g. Ag\'undez et al.~2008; Woods \& Willacy 2009).  Another disk scenario that may cause the extinction seen toward IRCs is if the disk around the primary extends into the line of sight to the IRC (e.g. VV CrA; Smith et al.~2009, Kruger et al.~2011).  Molecular absorption seen in this geometry would originate in the outer regions of the primary disk, likely resulting in largely different temperatures and column densities than expected toward an edge-on disk.  Thus it would be important to discern between these disk geometries before further studies of edge-on disks are confused by these disk systems.


DoAr 24Eb is a classical infrared companion to the young star DoAr 24Ea, seen with a separation of 2$^{\prime\prime}$.03 with the IRC at a position angle of 150$^{\circ}$ (E of N) relative to the primary (Ghez et al.~1993).  Koresko et al.~(1997) modeled the primary as being a K0 type star with temperature 5240 K, while the IRC was modeled as having a central stellar temperature of 4850 K.  We observed this system as part of our {\it Spitzer} Infrared Spectrograph (IRS) observing program of edge-on circumstellar disks and IRCs.  We present here findings for the DoAr 24E system and investigate the disk geometry of this system and the reason for the high extinction toward DoAr 24Eb.  In section 2, we introduce the observations taken.  In section 3, we describe the data reduction and discuss the detections.  We then discuss modeling and conclusions in section 4, and give a summary of our findings in section 5.

\begin{figure*}
\begin{center}
\includegraphics[clip,angle=90,scale=0.6]{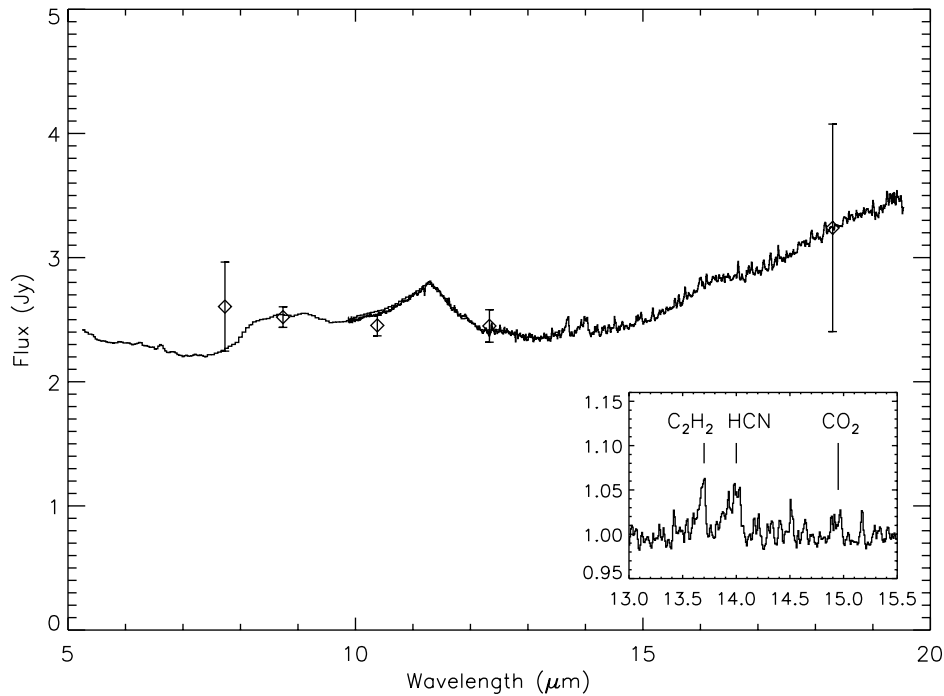}
\caption{{\it Spitzer} IRS spectrum of DoAr 24E, combining both binary components. Overplotted are the combined photometric fluxes measured from our T-ReCS imaging, summing 95\% of the flux from the DoAr 24E components.  The inset indicates the C$_2$H$_2$, HCN, and CO$_2$ emission features. }
\end{center}
\end{figure*}
\section{Observations}

\subsection{Spitzer IRS Spectrum}

DoAr 24E was part of our {\it Spitzer} Cycle 5 General Observers campaign,
Program ID 50152, to use the Infrared Spectrograph (IRS; Houck et al.~2004) in short-low and short-high (SL and SH) stare mode. 
The SL module has a slitwidth of 3.6$^{\prime\prime}$, plate scale of 1.8$^{\prime\prime}$ pixel$^{-1}$ (270 AU pixel$^{-1}$ assuming a distance of 150 pc to the system), and covers the spectral range 5.2-14.5 \textgreek{m}m with resolution R=60-127, while the SH module has a slitwidth of 4.7$^{\prime\prime}$, plate scale of 2.3$^{\prime\prime}$ pixel$^{-1}$ (350 AU pixel$^{-1}$), and covers 9.9-19.6 \textgreek{m}m with R=600.  
The
observations were designed to search for weak absorption features and 
so followed the observing technique described in Carr \& Najita (2008). We shortened the exposure time of each data frame to 6 seconds
and increased the number of integrations to verify repeatability of
the observations. Each setting (SL1, SL2, and SH) had twelve on-source
data frames at each nod position, accompanied with half as many off-source
integrations of the blank sky for hot pixel subtraction and to remove background emission.
The observation (AOR: 25679616) was taken on 2008 September 11 at 
$16^{h}26^{m}23.36^{s}\mbox{ }-24^{\circ}21^{\prime}1^{\prime\prime}.6$.
The coordinates were chosen to center the slit on DoAr 24Eb, calculated
by using the position of DoAr 24Ea from SIMBAD and the offsets found
by Chelli et al.~(1988). The observation used the nominal pointing
accuracy of $0^{\prime\prime}.1$.

The slit position angles during the observations were 139.4$^{\circ}$ for the SH module and 2.6$^{\circ}$ for the SL.  Given the primary-IRC position angle of 150$^{\circ}$ and separation 2$^{\prime\prime}$.03 (Ghez et al.~1993), and the widths of the SH and SL slits (4$^{\prime\prime}$.7 and 3$^{\prime\prime}$.6, respectively), we would expect a majority of the flux from both binary components to be within the slit for both observations.  A 4\% continuum flux difference between the SH and SL modules was corrected by scaling the SL spectrum by a constant, without any adjustments to the continuum shape, to match the SH spectrum at wavelengths where the orders overlap.

\subsection{Phoenix Spectrum}

On 2009 April 6, we took long-slit spectra with Phoenix (Hinkle et al.~2002) on Gemini South to examine the CO fundamental vibrational transitions. We used the standard
four pixel slit ($0.35^{\prime\prime}$) with R=50,000, along with
the M2150 filter with the echelle angle oriented to detect the $^{12}$CO
v=1-0 R(3)-R(5) transitions near 4.62 \textgreek{m}m, and the M2030 filter
for the P(20)-P(21) transitions near 4.86 \textgreek{m}m. Each ABBA nod position had a 120 second integration
with four 30 second exposures co-added internally. Standard star observations
were taken before the science frames for telluric division.
Only DoAr 24Eb was within the slit during the observation due to a sign error in the position angle given for the queue observation.

\subsection{T-ReCS Imaging}

To search for extended emission and to find flux ratios between binary
components, we used T-ReCS (Telesco et al.~1998) on Gemini South to
image DoAr 24E in 6 filters ranging from 7.7-18.3 \textgreek{m}m (see Table
1). We took calibration
images of standard stars before and after DoAr 24E.  Observations were taken by cycling through all six filters once on target. 
We used the standard 3 position chop-nod observing sequence and used the off-beams only for background subtraction.  These observations were taken 2009 June 8.

\begin{table}
\begin{center}
\caption{Fluxes as Measured with T-Recs Imaging}
\begin{tabular}{lcc}
\hline 
Filter & DoAr 24Ea (Jy) & DoAr 24Eb (Jy) \tabularnewline
\hline
\hline 
Si-1, 7.73 \textgreek{m}m & 0.79$\pm$0.2 & 1.96$\pm$0.3 \tabularnewline
Si-2, 8.74 \textgreek{m}m & 0.71$\pm$0.05 & 1.94$\pm$0.06\tabularnewline
Si-4, 10.38 \textgreek{m}m & 0.66$\pm$0.05 & 1.92$\pm$0.07\tabularnewline
Si-6, 12.33 \textgreek{m}m & 0.69$\pm$0.08 & 1.88$\pm$0.1\tabularnewline
Qa, 18.3 \textgreek{m}m & 1.16$\pm$0.5 & 2.25$\pm$0.7\tabularnewline
\hline
\end{tabular}
\end{center}
\end{table}

\section{Data Reduction and Detections}

\subsection{{\it Spitzer} IRS Spectrum}

For both the high resolution SH and low resolution SL spectra, we created custom IDL routines to identify and remove bad pixels on the array found using the images of the nearby blank sky.  The SL spectra were then reduced using optimal point source
extraction in the {\it Spitzer} IRS Custom Extractor (SPICE; version 2.2).
The high resolution SH spectra were reduced following the reduction
technique described in Carr \& Najita (2008) using custom IDL routines
along with standard spectral reduction routines in IRAF. 
We first used the \emph{apall} routine in IRAF to extract the raw spectra.
We used the standard star HR 6688 for flux calibration and to remove fringing.
Because fringing is dependent on the target position in the slit, we used
the observation of HR 6688 that resulted in the least amount of
fringing at shorter wavelengths after division.  We then used IRSFRINGE in IDL to remove any residual fringing.  The final spectrum is shown in Figure 1.

The {\it Spitzer} IRS spectrum shows warm molecular H$_2$O, C$_2$H$_2$, HCN, and CO$_2$ emission as was found by Pontoppidan et al.~(2010).  We examined the archival {\it Spitzer} IRS data reported by Pontoppidan et al.~(2010) and do not detect any significant changes in emission.  Further emission studies of DoAr 24E will be carried out with ground-based data by Moerchen et al.~({\it in prep.}).  The system also shows a double-peaked spectrum in the 8-12 \textgreek{m}m region.  The shape of this spectral region is similar to SVS20 (cf. Figure 7; Alexander et al.~2003), which was found to be due to hot, optically thin silicate emission being extinguished by cool, optically thin silicate absorption.  Such a combination of silicate emission and absorption is a common interpretation for  the relatively flat spectra found in YSOs (Mitchell \& Robinson 1981; Furlan et al.~2008).  Alternatively, the peaks at 9.2 and 11.3 \textgreek{m}m may be due to crystalline silicates which are also commonly seen in YSOs (Olofsson et al.~2009).  The lack of a strong silicate absorption feature is surprising given the previous high extinction estimates (i.e. 26 mag; Koresko et al.~1997).  Using the relation A$_{V}/\tau_{SiO}=16.6\pm2.1$ from Rieke \& Lebofsky (1985) for grains in the ISM, we would expect to see a silicate absorption feature with $\tau_{SiO}=1.6\pm0.2$.

\subsection{Phoenix Spectrum}

The near-IR Phoenix spectra of DoAr 24Eb were reduced in IRAF. We first used the
\emph{identify} task on the sky emission lines to get a wavelength
solution and to rectify the slit in each frame, with the rest frame
wavelengths of the emission lines taken from the HITRAN database (Rothman et al.~1998).

In the Phoenix spectrum, we found $^{12}$CO transitions both in emission and absorption, shown in Figure 2.  The emission component at least partially arises from warm gas as we see the v=2-1 R(11)-R(13) and P(15) transitions, as well as the high-J v=1-0 P(21) transition.  There are two discrete absorption features, found exclusively in the low-J R(3)-R(5) transitions, at v$_{LSR}=-5.8$ km s$^{-1}$ (hereafter A-6) and 3.2 km s$^{-1}$ (A3).  We fit the normalized, average profile of the R(3)-R(5) transitions using Gaussian distributions for both the emission and absorption features.  The fit is shown in Figure 3, and the Gaussian parameters used are given in Table 2.  In calculating the equivalent widths, given in Table 3, we assume the relevant local continuum includes the emission component as appropriate if the emission is coming from the innermost regions of the disk and the absorbing gas is further out.  We also assume the covering factor, the fraction of the unresolved emitting area that is being absorbed, to be 1.

\begin{figure*}
\begin{center}
\includegraphics[clip,angle=90,scale=0.7]{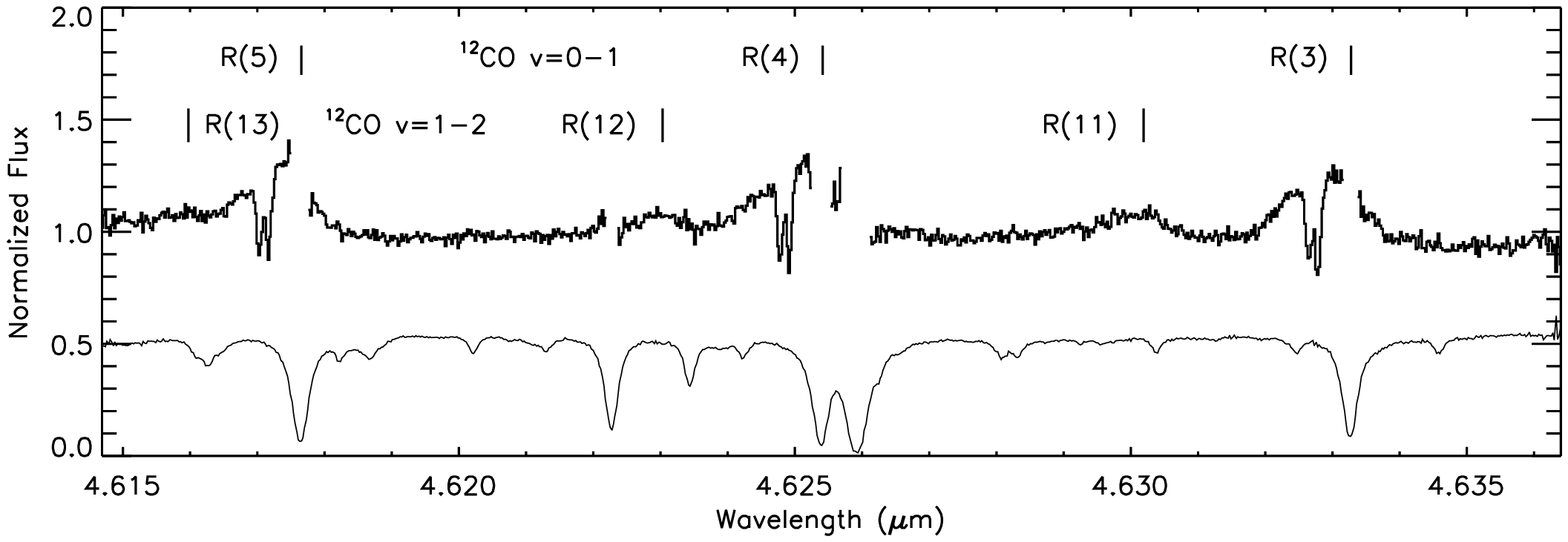}
\includegraphics[clip,angle=90,scale=0.7]{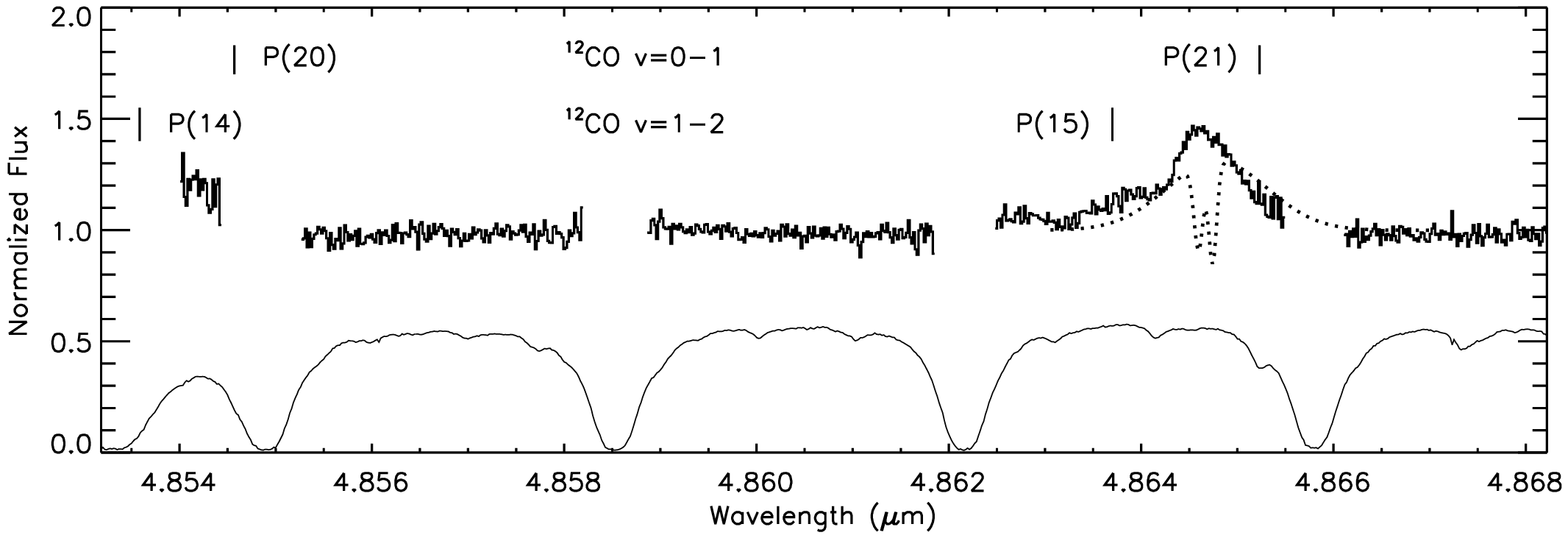}
\caption{CO fundamental spectra of DoAr 24Eb as observed by Phoenix in the topocentric frame in orders 16 (above) and 15 (below) with detected transitions marked.  The telluric transmission spectrum from a standard star is also plotted below each science spectrum.  The dotted line in order 15 is the average velocity profile fit to the R(3)-R(5) transitions, shown in Figure 3, centered at the P(21) transition for comparison.}
\end{center}
\end{figure*}


\begin{figure}
\includegraphics[clip,angle=90,scale=0.41]{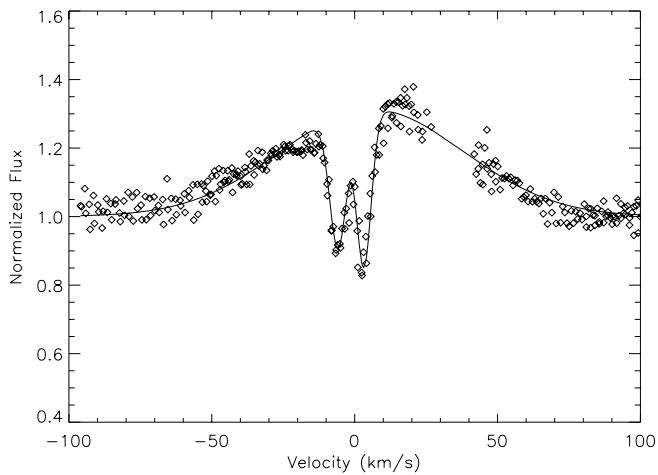}
\caption{The normalized, average CO line profile from the Phoenix using the R(3)-R(5) transitions.  The solid line is a fit using Gaussian profiles with the parameters listed in Table 2.  The missing data points are due to telluric absorption.}
\end{figure}

\begin{table}
\begin{center}
\caption{Measured Parameters for the $^{12}$CO Line Profiles}
\begin{tabular}{lcc}
\hline 
 & v$_{LSR}$ (km s$^{-1}$) & FWHM (km s$^{-1}$) \tabularnewline
\hline
\hline 
Emission & $6.1\pm0.6$ & $76\pm2$ \tabularnewline
Absorption (A-6) & $-5.8\pm0.2$ & $6.8\pm0.3$ \tabularnewline
Absorption (A3) & $3.2\pm0.3$ & $6.1\pm0.8$ \tabularnewline
\hline
\end{tabular}
\end{center}
\end{table}

\begin{table}
\begin{center}
\caption{Measured Equivalent Widths of the $^{12}$CO Fundamental Absorption
Lines}
\begin{tabular}{lcc}
\hline 
Component & Line ID & Equivalent Width (cm$^{-1}$)\tabularnewline
\hline
\hline 
A3 & R(3) & $0.0173\pm0.001$\tabularnewline
 & R(4) & $0.0178\pm0.001$\tabularnewline
 & R(5) & $0.0160\pm0.001$\tabularnewline
A-6 & R(3) & $0.0145\pm0.0009$\tabularnewline
 & R(4) & $0.0155\pm0.0009$\tabularnewline
 & R(5) & $0.0143\pm0.0009$\tabularnewline
\hline
\end{tabular}
\end{center}
\end{table}

\subsection{T-ReCS Imaging}

We used custom IDL routines to reduce the T-ReCS images. High winds
during the observation greatly reduced the image quality, so the images
were centered on the peak of the target flux before combining. We searched for extended emission by varying the extraction apertures, but
found none.  
Deconvolution using \emph{lucy} in IRAF gave the same results.  

We detected both binary components in all images.
We used a $1^{\prime\prime}$ aperture radius to estimate the photometric fluxes for the individual binary components, and the results are listed in Table 1.  We find a binary separation of $2^{\prime\prime}.03\pm0^{\prime\prime}.01$ at a position angle of $149^{\circ}.0\pm0^{\circ}.1$.  Ghez et al.~(1993) also measured a separation of $2^{\prime\prime}.03$ in July 1990, so we conclude that these objects have similar proper motions and are associated with each other.  As the change in separation since this observation is $<0^{\prime\prime}.01$, this shows the velocity difference between the objects in the direction perpendicular to our line of sight is <3.7 km s$^{-1}$, assuming the system is 150 pc away.  We also find the combined fluxes are similar to the {\it Spitzer} IRS spectrum, shown in Figure 1, assuming a 5\% flux loss from both components during the IRS observation.  We assume a flux loss from both components as the IRS slit was within 10$^{\circ}$ of the actual position angle of the DoAr 24E components during the observation.

\section{Discussion}

\subsection{CO Gas}
Najita et al.~(2003) showed that CO emission is commonly detected toward T Tauri stars, and most likely originates from the inner 5 AU of the disk.  They found the emission lines typically had FWHM of $\approx70$ km s$^{-1}$, similar to that found in the R(3)-R(5) transitions of DoAr 24Eb (FWHM$=76\pm2$ km s$^{-1}$).  We thus continue with the assumption that this emission largely originates from the inner 5 AU of the disk of DoAr 24Eb.  We do note, however, the P(21) transition has a different velocity profile than the R(3)-R(5) transitions as it peaks at -15 km s$^{-1}$, has a FWHM of 45 km s$^{-1}$, and is not Gaussian.  While the v=2-1 P(15) transition complicates the profile, the red side of the v=1-0 P(21) transition appears to be broader than the blue side.

We modeled the CO gas absorption components to extract temperature and column density estimates.  The velocity widths of these components are near the best-case instrument resolution of 5.88 km s$^{-1}$ (Hinkle 1999), so we cannot get precise measurements, but we can find a range of values that would result in the observed equivalent widths.  We modeled the transitions using a uniform temperature slab model with a covering factor of 1 and assuming the gas is in local thermodynamic equilibrium.  We considered a wide range of temperatures and column densities and used a $\chi^2$ goodness-of-fit to find the best-fit parameters.  The results are shown in Figure 4.

\begin{figure}
\includegraphics[clip,scale=0.4]{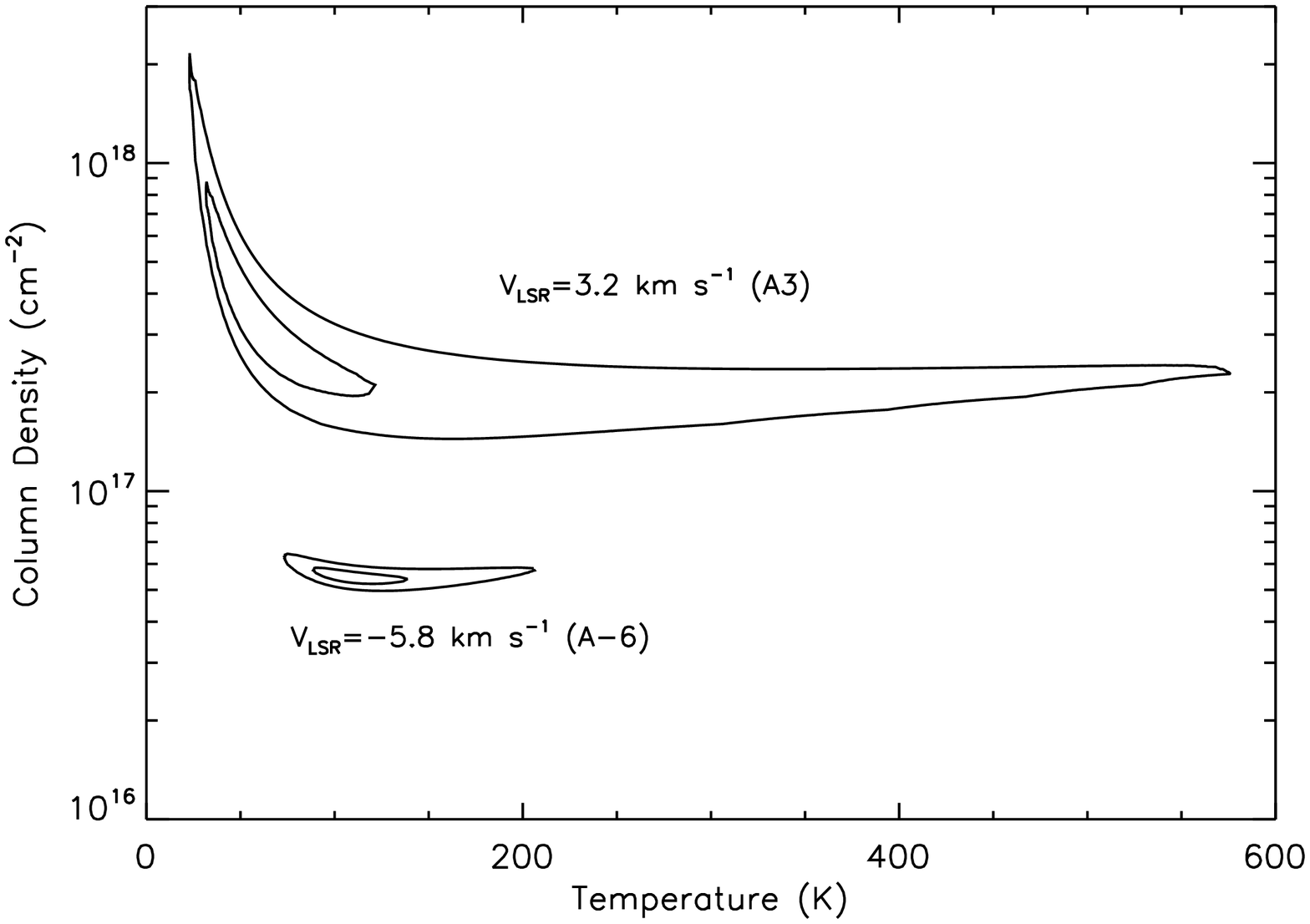}
\caption{The goodness-of-fit contours for the temperature and column density for the two absorption components, indicated by the velocity shifts of the corresponding components.  The contours show the 68\% and 90\% confidence levels assuming the best-case instrument resolution of 5.88 km s$^{-1}$ (see text).}
\end{figure}

The CO absorption feature at v$_{LSR}=3.2$ km s$^{-1}$ (A3) has a small FWHM of 6.1 km s$^{-1}$.  Using the best-case instrument resolution of 5.88 km s$^{-1}$, we modeled the absorbing gas with an intrinsic velocity width of 1.6 km s$^{-1}$.  The temperature and column density are not well constrained; the temperature we found is 75$\pm$50 K with a column density of $5.5\pm4.0\times10^{17}$ cm$^{-2}$, but these parameters are sensitive to the turbulent velocity.  If we assume an instrument resolution of 6.0 km s$^{-1}$, just less than the velocity width of A3 and corresponding to an intrinsic turbulent velocity of 1.1 km s$^{-1}$, the estimated 1-$\sigma$ column density increases to $2.5\pm1.8\times10^{19}$ cm$^{-1}$ and the temperature decreases to $33\pm10$ K.

The absorption feature at v$_{LSR}=-5.8$ km s$^{-1}$ (A-6) has a FWHM of 6.8 km s$^{-1}$.  We modeled these transitions with a turbulent velocity of 3.4 km s$^{-1}$ and found a best-fit temperature of $115\pm30$ K and column density $5.5\pm0.3\times10^{16}$ cm$^{-2}$.  These estimates show little dependence on the turbulent velocity used.  Again using an instrument resolution of 6.0 km s$^{-1}$ and resulting turbulent velocity of 3.2 km s$^{-1}$, we find a temperature estimate of $105\pm20$ K and column density $5.9\pm0.3\times10^{16}$ cm$^{-2}$.  

To further investigate the CO absorption, we obtained spectroscopic data from the CRyogenic high-resolution InfraRed Echelle Spectrograph (CRIRES; K\"aufl et al.~2004) that was reported in Pontoppidan et al.~(2011).  These spectra include the $^{12}$CO R(0)-R(1) and P(1)-P(9) transitions and were observed on 2007 September 3.  The spectrum of DoAr 24Ea shows only the A3 absorption lines.  The emission line profile for DoAr 24Eb does not appear to change between the low- and high-J transitions, up to P(9).  We also found the A3 transitions have an average FWHM width of 3.6 km s$^{-1}$, while the A-6 lines have an average width of 4.9 km s$^{-1}$.  Given the measured instrument resolution of 3.18 km s$^{-1}$ (CRIRES User Manual, Issue 85.2), this would indicate FWHM turbulent velocity widths of 1.7 and 3.7 km s$^{-1}$, respectively.  This is comparable to the turbulent velocity estimates with the Phoenix data (1.6 and 3.4 km s$^{-1}$, respectively).  The A3 line is only seen up to the P(7) transition, showing it originates in cooler gas (<50 K).  It is difficult to estimate the optical depth of the A-6 transitions due to being so near the strong telluric CO absorption and the emission line profile changing for higher transitions.  However, the A-6 absorption lines are seen through the P(9) transition, and we find that modeling the gas with temperature 115 K and column density $5.5\times10^{16}$ cm$^{-2}$, as found with the Phoenix spectrum, fits the CRIRES spectrum well.

The emission line in the CRIRES spectrum is slightly narrower than found with Phoenix, and peaks at negative velocities.  If we again estimate the emission as a Gaussian, we find it has a FWHM of 54 km s$^{-1}$ and v$_{LSR}=-9$ km s$^{-1}$ (see Figure 5).  This change in velocity from the emission seen in the Phoenix spectrum may be due to complicated details within the inner regions of the disk, possibly due to DoAr 2Eb itself being a binary (Koresko 2002).

\begin{figure}
\includegraphics[clip,angle=90,scale=0.41]{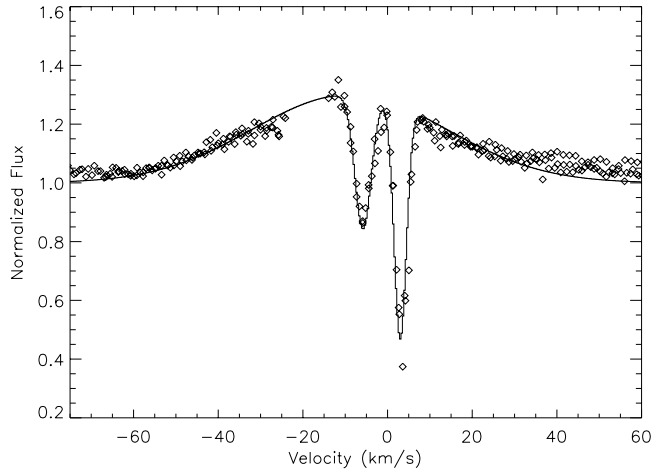}
\caption{The normalized, average CO line profile from the CRIRES spectrum using the P(3)-P(5) transitions.  The solid line is a fit to the data with the emission estimated as a Gaussian with FWHM=54 km s$^{-1}$ and v$_{LSR}=-9$ km s$^{-1}$.  The two absorption lines were found by using the turbulent velocity estimates and average absorption strengths found in the Phoenix spectrum, convolved with the CRIRES resolving power of 3.18 km s$^{-1}$ (see text), showing a good fit to the absorption lines in the CRIRES spectrum.  }
\end{figure}

\subsection{Gas Location}

There are several possible locations for the two absorbing CO gas columns, among which are a disk, disk wind, or envelope associated with DoAr 24Eb, the outer regions of a disk around DoAr 24Ea, or the interstellar medium (ISM).

\begin{figure*}
\begin{center}
\includegraphics[scale=0.5]{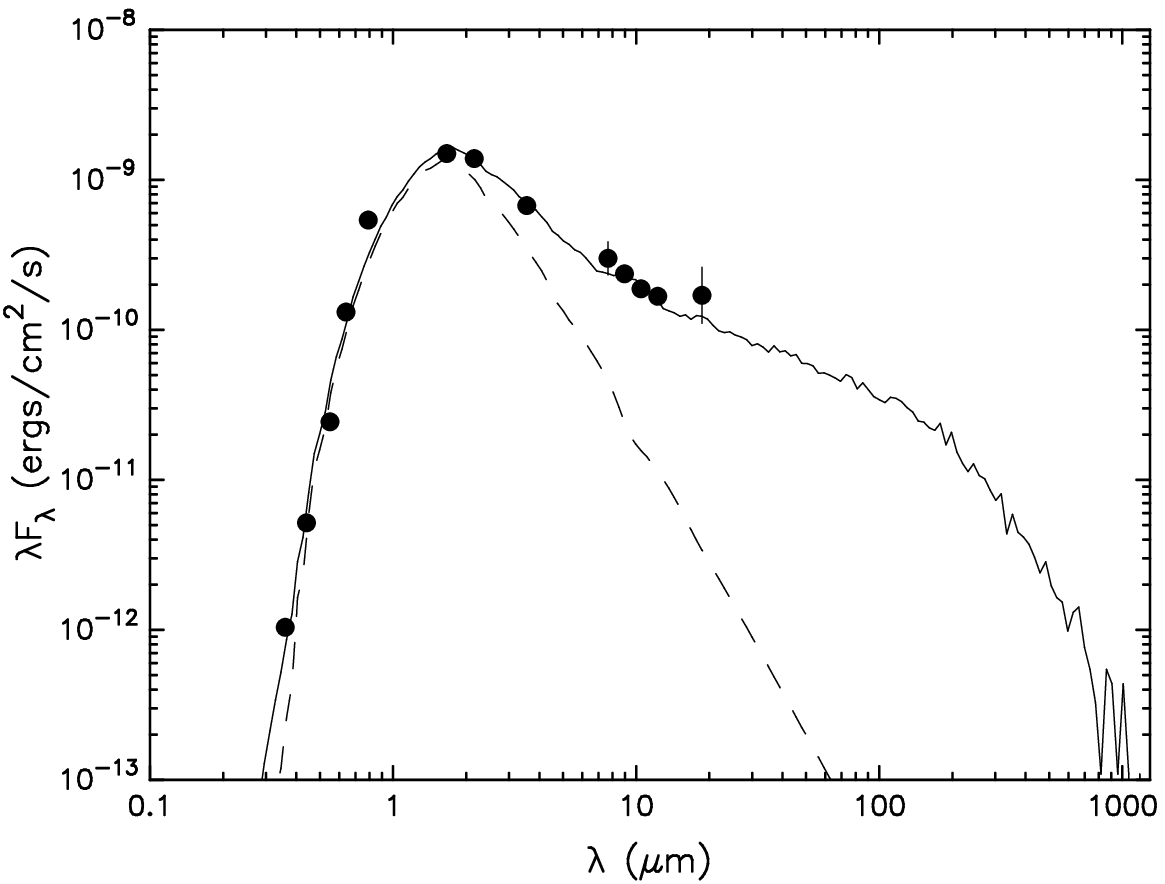}
\includegraphics[scale=0.5]{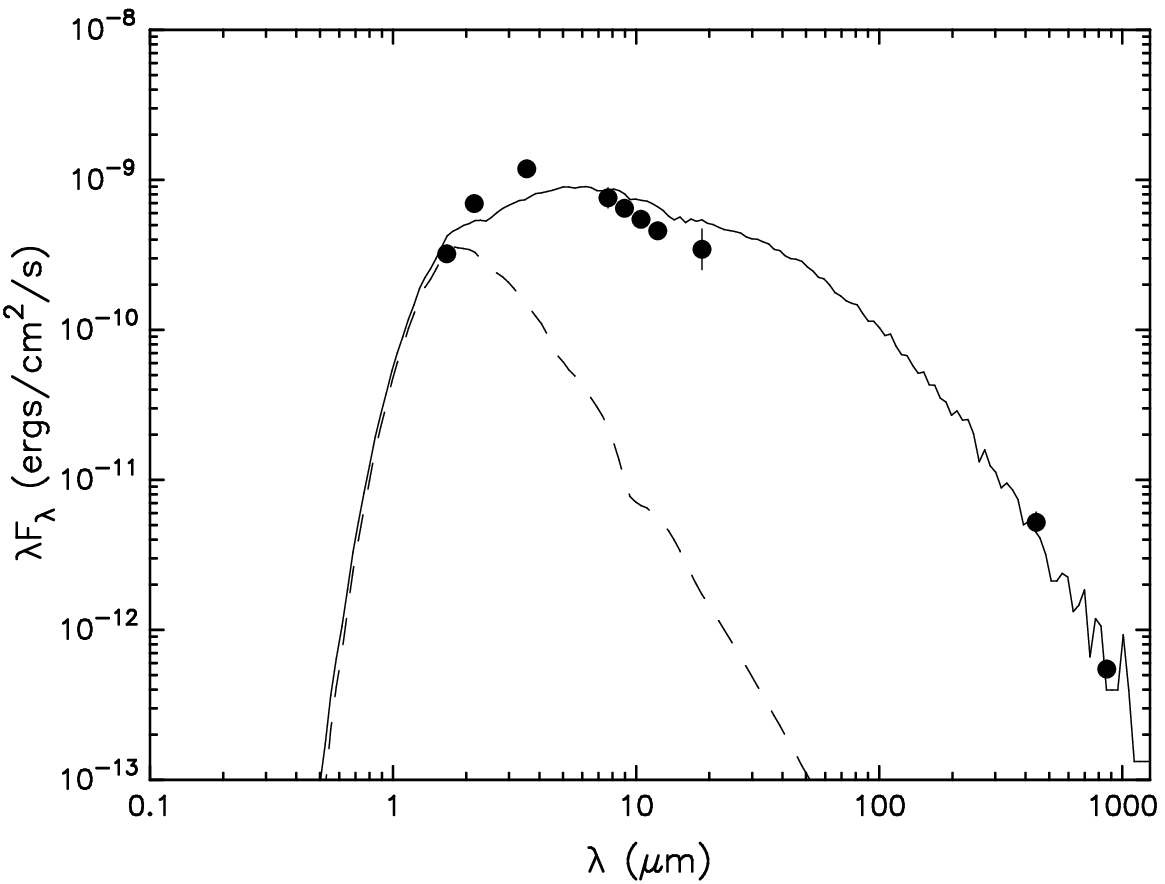}
\includegraphics[scale=0.5]{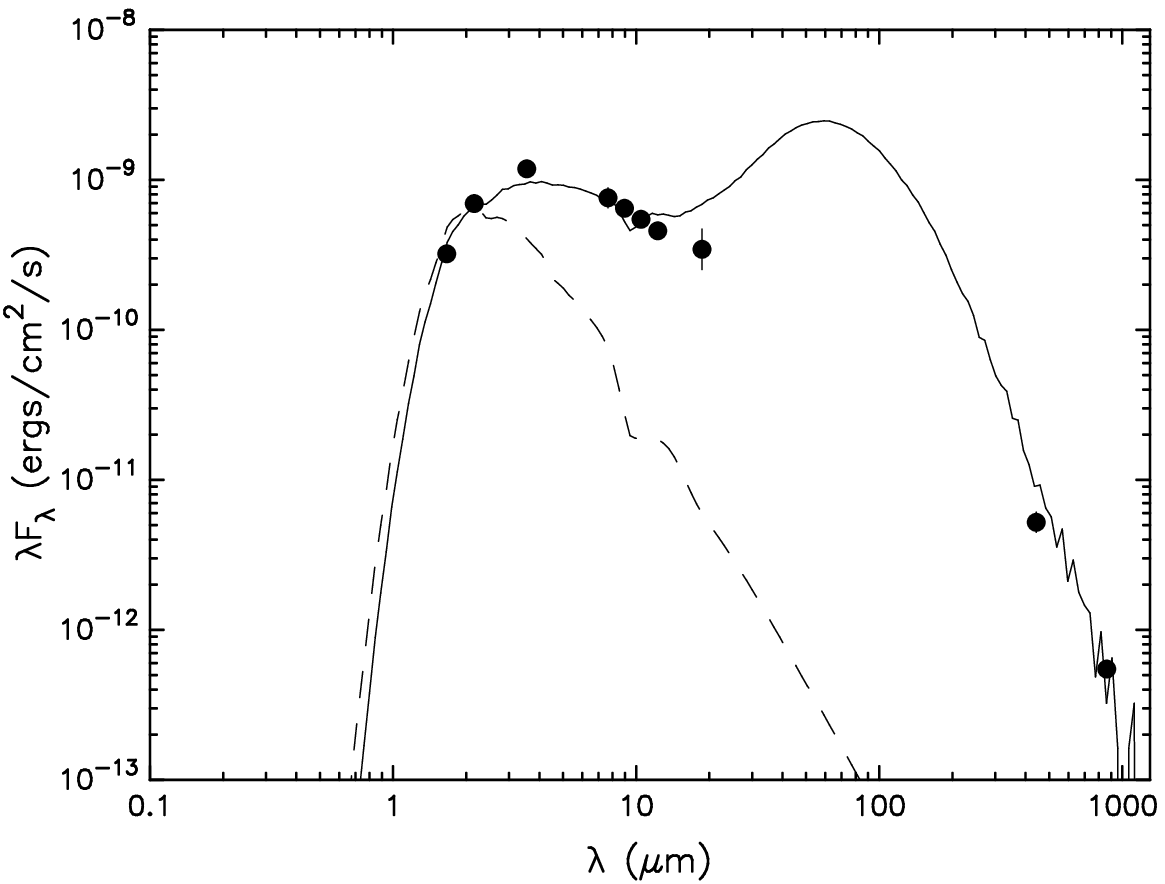}
\caption{The best-fit models to the SEDs of DoAr 24Ea and DoAr 24Eb.  The dots are spectral points taken from our T-ReCS imaging and Chelli et al.~(1988).  The solid lines are the SED fits, while the dashed lines are the stellar photospheres as it would look without a circumstellar disk but still including the interstellar extinction.  Left: The spectrum of DoAr 24Ea.  Middle: The spectrum of DoAr 24Eb modeled as having an interstellar extinction <10 mag.  Right: The spectrum of DoAr 24Eb modeled as having an interstellar extinction <30 mag.   }
\end{center}
\end{figure*}

The main source of interstellar extinction is likely the $\rho$ Ophiuchus ($\rho$ Oph) cloud.  This cloud has a system velocity of v$_{LSR}\approx3-4$ km s$^{-1}$, similar to the A3 CO.  Chelli et al.~(1988) estimated an interstellar extinction of A$_{V}=6.2$ mag for DoAr 24Ea, so if we use the interstellar ratio of N(CO)/A$_{V}=1.4\times10^{17}$ cm$^{-2}$ mag$^{-1}$ (Rettig et al.~2006), we would expect a CO column density of $8.7\times10^{17}$ cm$^{-1}$.  This column density is within the 1-$\sigma$ error of the estimate for the A3 CO, and the low temperature of the absorbing gas (<50 K) is typical for the ISM.  The CRIRES spectrum also includes DoAr 24Ea within the slit (although the spectrum of DoAr 24Ea was not included in Pontoppidan et al.~2011) and we find that DoAr 24Ea also shows the A3 line with a similar velocity profile and equivalent width as DoAr 24Eb.  We conclude the A3 CO gas is associated with the ISM.

The A-6 line is not seen toward DoAr 24Ea in the CRIRES spectrum, so this CO is associated with the material that is extinguishing DoAr 24Eb but not DoAr 24Ea.  Koresko et al.~(1997) used H and K band photometry to model the infrared companion DoAr 24Eb and estimated a high extinction of A$_{V}=26$ mag.  Removing the interstellar extinction, this would indicate A$_{V}\approx20$ mag for the material that is obscuring DoAr 24Eb but not DoAr 24Ea.  Again comparing to the N(CO)/A$_{V}$ ratio of the ISM, this would indicate a high CO column density of $2.8\times10^{18}$ cm$^{-2}$, which is nearly 2 orders of magnitude greater than what we estimated for the A-6 CO.  While this implies a substantially smaller gas/dust ratio than is seen in the ISM ($\approx1.5$\%), it may be because the M-band continuum is coming from a larger region of the disk than where the near-infrared extinction is occurring, resulting in a covering factor of <1.  This may also explain why we do not see a large silicate absorption feature in the {\it Spitzer} IRS spectrum as expected for a system showing such high extinction.

While the location of the A-6 CO is unclear, it would not be associated with the outer disk regions of DoAr 24Ea.  Assuming a distance of 140-160 pc to the system, the binary components have a projected separation of 285-325 AU, and the gas temperature of $\sim100$ K is high for these disk radii. Furthermore, Doppmann et al.~(2003) observed the photospheric absorption of DoAr 24Ea (GY 20A) to find a redshift of v$_{LSR}=6.4$ km s$^{-1}$.  If the A-6 CO was associated with DoAr 24Ea, the gas would have a 12.2 km s$^{-1}$ velocity shift from the central stellar component, much too high for gas at these disk radii.  With such a large column only seen toward DoAr 24Eb, and having a discrete velocity shift of -9 km s$^{-1}$ from the surrounding cloud, it is difficult to envision a physical situation in which this is not associated with the disk of DoAr 24Eb.  It is unknown if the A-6 line is at the system velocity of DoAr 24Eb, or possibly shifted from the central star velocity, such as in a disk wind.

\subsection{SED Modeling}

To further investigate the system, we used the spectral energy distribution (SED) model fitter described in Robitaille et al.~(2006).  This tool finds a best-fit to an input SED using a database of 200,000 pre-computed radiation transfer models of circumstellar disks, with 14 physical parameters varied at 10 different disk viewing angles ranging from 18.2 to 87.1$^\circ$.  The models differentiate the circumstellar extinction, due to the disk material between the stellar surface and the outer edge of the disk, and the interstellar extinction between the disk edge and the observer.  The user inputs spectral points, and can constrain the distance to the system and the interstellar extinction.  The best-fit models are then returned in order of increasing $\chi^2$ value.  The estimates for the distance to the $\rho$ Oph star forming region are in the 140-160 pc range, we found that using this limited distance range resulted in only a few returned models.  We thus extended the distance range by 20 pc in both directions (120-180 pc) in order to get a range of disk parameters.  We then used only those models with $\chi_{\mbox{best}}^2-\chi^2<5\mbox{N}$, where N is the number of input data points, and stellar temperatures within 500 K of estimates in Koresko et al.~(1997).  We show the best-fit models described below in Figure 6.  Table 4 reports parameters for the best-fit models, as well as the minimum and maximum values of each individual parameter for the models returned given the $\chi^2$ cut-off.

We first fit the SED of DoAr 24Ea for comparison with predictions.  We constrained the interstellar extinction to be in the 0-10 mag range, and find the models are close to previous estimates.   While Koresko et al.~(1997) modeled this system as having a central stellar temperature of 5240 K, the model stellar temperatures are consistently lower ($\approx4800$ K).  This may account for the modeled interstellar extinction (5.3 mag) being slightly lower than previous estimates (e.g. 6.2 mag; Chelli et al.~1988).  The models predict the system does not have an infalling envelope.

\begin{table*}
\begin{center}
\caption{Estimated Disk Parameters from SED Fitting}
\begin{tabular}{lccc}
\hline
& & DoAr 24Eb & DoAr 24Eb\tabularnewline
Parameter & DoAr 24Ea & (A$_V=0-10$ mag)$^{a}$ & (A$_V=0-30$ mag)$^{a}$\tabularnewline
\hline
\hline
Interstellar A$_V$ (mag) & 5.3 (5.2 5.3) & 10 (1.6 10) & 17 (11 24) \tabularnewline
Circumstellar A$_V$ (mag) & 2.2e-4 (2.2e-4 6.8e-4) & 0.026 (0.026 150)$^{b}$ & 2.5 (0.0007 4.6)  \tabularnewline
Stellar Temperature (K) & 4810 (4710 4810) & 4550 (4400 5100) & 4600 (4400 5100) \tabularnewline
Stellar Mass (M$_{\odot}$) & 2.0 (1.8 2.0) & 1.5 (1.3 3.1) &  1.9 (1.5 3.5) \tabularnewline
Env. Infall Rate ($10^{-6}$ M$_{\odot}$ yr$^{-1}$) & 0 (0 0) & 5.5e-3 (0 13) & 2.1 (0 3.8) \tabularnewline
Disk Accretion Rate ($10^{-8}$ M$_{\odot}$ yr$^{-1}$) & 8.3 (8.3 11) & 86 (0.26 86) & 6.1 (6.3e-3 86)  \tabularnewline
Disk Inner Radius (AU) & 0.15 (0.15 0.15) & 1.4 (0.21 1.6) & 0.63 (0.26 2.3) \tabularnewline
Cavity Opening Angle ( $^{\circ}$ ) & No env. & 32 (19 40)$^{c}$ & 35 (32 54)$^{c}$ \tabularnewline
Disk Inclination ( $^{\circ}$ ) &76 (63 81) & 76 (63 81) & 70 (32 81) \tabularnewline
\hline
\\
\multicolumn{4}{l}{{\bf Notes.} Shown are the best-fit and the (min, max) of the returned parameters for the best-fit models.}\tabularnewline
\multicolumn{4}{l}{$^{a}$Interstellar extinction constraint.}\tabularnewline
\multicolumn{4}{l}{$^{b}$Median value of 12 mag, see text.} \tabularnewline
\multicolumn{4}{l}{$^{c}$The range for those disks with an envelope.}\tabularnewline

\end{tabular}

\end{center}

\end{table*}

To investigate the scenario where the extinction in DoAr 24Eb is due to a circumstellar disk, likely seen near edge-on, we again constrained the interstellar extinction to be in the 0-10 mag range.   The returned models have interstellar extinctions that are higher than predictions, tending to be at the upper limit of 10 mag.  The best-fit model also surprisingly predicts a small circumstellar extinction A$_V$ of only 0.026 mag.  The extinction estimates go as high as 150 mag, but the median extinction estimate in the returned models is 12 mag, closer to expectations given interstellar extinction constraint and previous total extinction estimates (i.e. 26 mag; Koresko et al.~1997).  Unlike DoAr 24Ea, most of these models also predict an infalling envelope.  Given that DoAr 24Eb is physically associated with DoAr 24Ea, as shown by the common proper motions, we would expect the systems to be the same age and likely the same stage of evolution.  Thus, having an envelope around DoAr 24Eb and not DoAr 24Eb would require some mechanism to either prolong the envelope around DoAr 24Eb or shorten the envelope around DoAr 24Ea.  On a cautionary note, the SED model fitter only tests disk inclinations at discrete intervals, which is a drawback for edge-on disks where small inclination changes result in large SED changes (Robitaille et al.~2007; Kruger et al.~2011).  Presumably, finer gradations in disk inclination would provide models with more appropriate circumstellar extinction estimates and possibly eliminate the envelopes.

For completeness, we modeled DoAr 24Eb as having another compact, interstellar component in the line of sight, as would be the case if the outer disk of DoAr 24Ea was in the line of sight, by allowing the interstellar extinction to be in the range 0-30 mag.  Because there were so many best-fit models returned, we lowered the $\chi^2$ cut-off to $\chi_{\mbox{best}}^2-\chi^2<3\mbox{N}$ so that we were not over-estimating our parameters (Robitaille et al.~2007).  The stellar temperatures are comparable to Koresko et al.~(1997), ranging above and below, and most models do not include a surrounding envelope.  The modeled interstellar extinctions are slightly lower but similar to previous extinction estimates for DoAr 24Eb (i.e. 26 mag; Koresko et al.~1997).  The models associated a majority of the extinction with the ISM, with the largest circumstellar extinction being 0.44 mag.  As discussed in section 4.2, it is unlikely that such a large column can be seen toward DoAr 24Eb, and not DoAr 24Ea, without being a part of the disk of DoAr 24Eb.

\section{Summary}

We investigated the source of the high extinction toward the infrared companion DoAr 24Eb, and considered several disk and envelope geometries that may explain the high extinction toward DoAr 24Eb:  a disk or envelope around DoAr 24Eb, the outer regions of a disk around DoAr 24Ea, or the interstellar medium (ISM).  To test the scenarios, we analyzed near-infrared spectra from Phoenix of the CO fundamental transitions and modeled the mid-infrared photometric fluxes we found with T-ReCS imaging.

DoAr 24Eb shows two $^{12}$CO absorption features at v$_{LSR}=3.2$ (A3) and -5.8 km s$^{-1}$ (A-6).  Using a slab model and optically thick gas, we found the A-6 CO has a temperature of $\sim100$ K and column density  $\sim6\times10^{16}$ cm$^{-2}$ while the A3 CO has a higher column density ($10^{17}-10^{18}$ cm$^{-2}$).  A CRIRES spectrum also indicates the A3 line has a low temperature (<50 K).  The velocity shift, column density, and temperature of the A3 line, as well as its presence in the DoAr 24Ea spectrum, indicate that it is associated with the interstellar medium.  

Since A-6 is seen toward DoAr 24Eb and not DoAr 24Ea, it is most likely associated with DoAr 24Eb.  This CO is blueshifted $\approx12$ km s$^{-1}$ from the radial velocity of DoAr 24Ea (Doppmann et al.~2003) and has a temperature of $\sim100$ K, so it would not be in outer disk regions of DoAr 24Ea in the line of sight to DoAr 24Eb.  The discrete shift from the cloud velocity and having such a large column in the line of sight toward only DoAr 24Eb make it most likely in a disk or disk wind associated with DoAr 24Eb.  

The {\it Spitzer} IRS spectrum does not show strong silicate absorption as we would expect for an object showing high extinction.  Also, our measurement of the column density for the A-6 CO absorption, which is associated with DoAr 24Eb, is two orders of magnitude smaller than expected for the high extinction.  These factors suggest the mid-infrared continuum is not as extinguished as the near-infrared.  This may be due to the source of the mid-infrared flux being more extended than the near-infrared, possibly due to an extended disk.  This further argues against the source of extinction toward the IRC being due to the disk around the primary or the ISM, which would be expected to show a more uniform extinction.

To further investigate the disk scenarios, we modeled the spectral energy distributions of both components.  We tested if the extinguishing material is likely associated with the disk around DoAr 24Eb by constraining the interstellar medium to be less that 10 mag, as is found for DoAr 24Ea.  We found the models predict a wide range of circumstellar extinctions and an infalling envelope for DoAr 24Eb, which is not predicted for DoAr 24Ea.  However, this may be due to the models only being tested at discrete intervals (Robitaille et al.~2007), which results in large SED changes for high inclinations.

\acknowledgements{

We would like to thank the anonymous reviewer
for the constructive comments and suggestions.  Support for this work was provided by the National Science Foundation under Grant No. AST-0708074, and by NASA through contract RSA No. 1346810, issued by JPL/Caltech. This work is based on observations made with the {\it {\it Spitzer} Space Telescope} and Gemini Observatory. The {\it {\it Spitzer} Space Telescope} is operated by the Jet Propulsion Laboratory, California Institute of Technology under a contract with NASA. The Gemini Observatory is operated by the Association of Universities for Research in Astronomy, Inc., under a cooperative agreement with the NSF on behalf of the Gemini partnership: the National Science Foundation (United States), the Science and Technology Facilities Council (United Kingdom), the National Research Council (Canada), CONICYT (Chile), the Australian Research Council (Australia), Minist\'{e}rio da Ci\^{e}ncia e Tecnologia (Brazil) and Ministerio de Ciencia, Tecnolog\'{i}a e Innovaci\'{o}n Productiva (Argentina). This work was also based on observations made with ESO Telescopes at the La Silla Paranal Observatory under programme ID 179.C-0151, and made use of the SIMBAD database, operated at CDS, Strasbourg, France.
}

\end{document}